\documentclass[english,12pt]{article}
\usepackage{array}
\usepackage{graphicx}
\usepackage{amssymb}
\usepackage[english]{babel}
\usepackage{amsmath}
\usepackage{multirow}
\usepackage{prettyref}
\usepackage{babel}
\usepackage{units}
\usepackage[latin1]{inputenc}
\usepackage{amsfonts}
\usepackage{amssymb}
\usepackage{babel}
\usepackage{color}
\usepackage{cite}
\def\@fmsl@sh#1#2#3{\m@th\ooalign{$\hfil#1\mkern#2/\hfil$\crcr$#1#3$}}
 \def\eq#1\en{\begin{equation}#1\end{equation}}
\def\s[#1,#2]{[#1\stackrel{\star}{,}#2]}
\def\sx[#1,#2]{[#1\stackrel{\star_{x}}{,}#2]}

\providecommand{\bra}[1]{\langle#1|}

\providecommand{\ket}[1]{|#1\rangle}

\def\bc{\begin{center}}
\def\ec{\end{center}}

\def\gsim{\mathrel{\mathpalette\atversim>}}

\def\bc{\begin{center}}
\def\ec{\end{center}}

\def\gsim{\mathrel{\rlap{\lower4pt\hbox{\hskip1pt$\sim$}}

    \raise1pt\hbox{$>$}}}       %greater than or approx. symbol

\def\gsim{\mathrel{\rlap{\lower4pt\hbox{\hskip1pt$\sim$}}
    \raise1pt\hbox{$>$}}}       %greater than or approx. symbol

\begin{document}
\makeatletter
\def\fmslash{\@ifnextchar[{\fmsl@sh}{\fmsl@sh[0mu]}}
\def\fmsl@sh[#1]#2{%
  \mathchoice
    {\@fmsl@sh\displaystyle{#1}{#2}}%
    {\@fmsl@sh\textstyle{#1}{#2}}%
    {\@fmsl@sh\scriptstyle{#1}{#2}}%
    {\@fmsl@sh\scriptscriptstyle{#1}{#2}}}
\def\@fmsl@sh#1#2#3{\m@th\ooalign{$\hfil#1\mkern#2/\hfil$\crcr$#1#3$}}
\makeatother
%\baselineskip 24pt

%%%%%%%%%%%%%%%%%%%%%%%%%%%%%%%%%%%%%%%%%%%%%%%%%%%%%%%%%%%%%%%%%
%%%
%%%                      TITLE PAGE
%%%
%%%%%%%%%%%%%%%%%%%%%%%%%%%%%%%%%%%%%%%%%%%%%%%%%%%%%%%%%%%%%%%%%
\thispagestyle{empty}
\begin{titlepage}
\boldmath
\begin{center}
  \Large {\bf Decoherence and Quantum Measurement: \\
  The Missing Lecture}
    \end{center}
\unboldmath
\vspace{0.2cm}

\begin{center}
{\large  Stephen D. H. Hsu}\footnote{hsusteve@gmail.com}
\end{center}

\begin{center}
{\sl Department of Physics and Astronomy\\ Michigan State University, East Lansing, Michigan 48823, USA
}\\

\end{center}
\vspace{2cm}
\begin{abstract}
\noindent
We give an elementary account of quantum measurement and related topics from the modern perspective of decoherence. The discussion should be comprehensible to students who have completed a basic course in quantum mechanics with exposure to concepts such as Hilbert space, density matrices, and von Neumann projection (``wavefunction collapse''). 
\end{abstract}
\vspace{5cm}
\end{titlepage}

\section{Introduction}

Decoherence and quantum measurement are usually neglected in standard courses on quantum mechanics. While the {\it Shut up and Calculate!} perspective is defensible in an introductory course, students are ultimately entitled to deeper explanations. We hope these notes are the equivalent of a Missing Lecture on what has become the widely accepted modern perspective regarding these topics. The notes are based on material covered in Physics 905 Advanced Quantum Mechanics, Michigan State University, 2022.

Quantum mechanics, as conventionally formulated, has two types of time evolution. 

\noindent 1. An isolated system evolves
according to the Schr\"{o}dinger equation
\begin{equation}
i \frac{\partial}{\partial t} \Psi ~=~  H \Psi~~.    
\end{equation}
The solution satisfies 
$\Psi (t) = U(t) \Psi (0)~$,
and the time evolution operator $U(t) = \exp (-iHt)$ is unitary: $U^\dagger U = 1$.

\noindent 2. However, when a measurement is made the state undergoes non-unitary von Neumann projection to an eigenstate corresponding to the observed eigenvalue. 
\smallskip

Because the two types of time evolution are so radically different, students typically demand (and indeed are entitled to demand) a rigorous definition of exactly when each of them apply. Under what conditions, exactly, does von Neumann projection occur? When and why does the system deviate from ordinary Schr\"{o}dinger evolution?

It is widely acknowledged that the conventional interpretation does not supply a satisfactory definition for when von Neumann projection (or ``wavefunction collapse'') applies. See, e.g., Against Measurement by J.S. Bell \cite{Bell}. 

The modern formulation of quantum mechanics does not require non-unitary wavefunction collapse. The phenomenology of von Neumann projection can be reproduced under unitary evolution of the system as a whole, due to a phenomenon known as {\it decoherence} \cite{deco1,deco2,deco3,deco4}, which we explain below. What is conventionally referred to as measurement is actually a continuous process that results from entanglement between the measured object and the many degrees of freedom in the measuring device or local environment.  

%A brief review of density matrices and reduced density matrics is given in the Appendix.

\section{Idealized Measurement of a Qubit}

\begin{figure}[htp]
    \centering
    \includegraphics[width=0.6\linewidth]{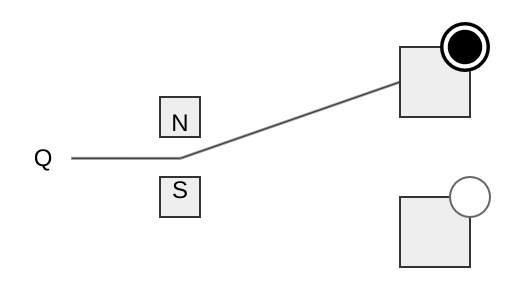}
    \caption{Idealized measurement of a single qubit (spin): $\ket{\pm}$ states are deflected up or down, entering one of the two detectors. Each detector has an indicator light which releases a macroscopic number of photons when activated. In the figure, a $\ket{+}$ spin is deflected upwards and enters the top detector, causing its indicator light to emit photons. The resulting state of the system is $\ket{S_+}$. A $\ket{-}$ spin state results in the system state $\ket{S_-}$ (not shown), in which the photons are emitted from the bottom detector. The states $\ket{S_+}$ and $\ket{S_-}$ differ macroscopically and have almost zero overlap.
    }
    \label{fig1}
\end{figure}

Let $Q$ be a single qubit and $M$ a macroscopic device which measures the spin of the qubit along a particular axis. The eigenstates of spin along this axis are denoted $\vert \pm \rangle$. We define the operation of $M$ as follows, where the combined system is $S = Q + M$. Note $S$ includes both the measuring device and the qubit $Q$.
\begin{align}
&\ket{+}\otimes\ket{M} \longrightarrow \ket{S_{+}}, \\
&\ket{-}\otimes\ket{M} \longrightarrow \ket{S_{-}},
\end{align}
where $S_+$ denotes a state of the total system in which the apparatus $M$ has recorded a $+$ outcome (i.e., is in state $M_+$), and similarly with $S_-$. 
See figure (\ref{fig1}).
We can then ask what happens to a superposition state 
$\ket{\Psi_Q} = c_+ \ket{+} + c_- \ket{-}$ which enters the device $M$. In the conventional formulation, with measurement collapse, {\it one} of the two final states $\ket{S_+}$ {\it or} $\, \ket{S_-}$ is realized, with probabilities $\vert c_+ \vert^2$ and $\vert c_- \vert^2$ respectively. 

However, if the combined system $S = Q + M$ evolves according to the Schr\"{o}dinger equation (in particular, linearly), we obtain a superposition of measurement device states:
\begin{align}
\bigl(  c_+ \ket{+} ~+~ c_- \ket{-}  \bigr) \otimes \ket{M} ~~ \longrightarrow ~~
c_+ \, \ket{S_+} ~+~ c_- \, \ket{S_-}.
\label{super}
\end{align}
This seems counter to actual experience: measurements produce a single outcome, not a superposition state. 

However, it is almost impossible for an observer in the state $S_+$ to be aware of the second branch of the wave function in state $S_-$. Any object sufficiently complex to be considered either a measuring device or observer (for example, which can be regarded as semi-classical) will have many degrees of freedom. A measurement can only be said to have occurred if the states $M_+$ and $M_-$ are very different: the outcome of the measurement must be stored in a redundant and macroscopically accessible way in the device (or, equivalently, in the local environment). In figure (\ref{fig1}), $M_+$ corresponds to the activation of the top detector, and a flash of its light. $M_-$ corresponds to the activation of the bottom detector and its light. Typically, the overlap of $M_+$ with $M_-$ is effectively zero: of order $\exp( - N )$, where $N$ is a macroscopic number of degrees of freedom. This tiny overlap generally persists under further dynamical evolution \cite{Buniy:2020dux}. 

For our purposes the phenomenology described above can be taken as a definition of what we mean by decoherence: two components of a given superposition state interact with, and become entangled with, environmental or measuring device degrees of freedom. The resulting environmental components of the superposition state are radically different and have almost zero overlap.

For All Practical Purposes, to use Bell's terminology \cite{Bell}, an observer on one branch can ignore the existence of the other: they are said to have decohered. Each of the two observers will perceive a collapse to have occurred, although the evolution of the overall system $S$ has continued to obey the Schr\"{o}dinger equation.

%Simple set-up as described in Steve's slides:
%\begin{equation}
%\vert \pm \rangle \otimes \vert M_0 \rangle ~\rightarrow~ {\bf SG} ~\rightarrow~ \vert \pm, M_\pm \rangle    
%\end{equation}

\bigskip

We can formalize this analysis using a density matrix. Taking $c_\pm$ equal for simplicity, the post-measurement state is
\begin{equation}
\vert \Psi \rangle = \frac{1}{\sqrt{2}} \, \Bigl( \vert +, M_+ \rangle ~+~ \vert -, M_- \rangle \Bigr)
\end{equation}
\begin{equation}
\rho = \vert \Psi \rangle \langle \Psi \vert    ~~.
\end{equation}
We obtain a reduced density matrix $\rho_Q$, describing only the qubit degrees of freedom, by tracing over the $M$ degrees of freedom. See Appendix for more details.
\begin{equation}
\rho_Q = \frac{1}{2} \, \Bigl( \vert + \rangle \langle + \vert  ~+~   \vert - \rangle \langle - \vert ~+~ 
\langle M_- \vert M_+ \rangle \, \vert + \rangle \langle - \vert ~+~ 
\langle M_+ \vert M_- \rangle \, \vert - \rangle \langle + \vert
\Bigr)   
\end{equation}
In a properly designed measurement, the states $M_\pm$ are macroscopically distinct, with very small overlap. 
Therefore, we have
\begin{equation}
\rho_Q \approx \begin{bmatrix}   1/2 & 0 \\ 0 & 1/2  \end{bmatrix}    
\end{equation}
which describes a mixed state. A mixed state is not a superposition, but rather can be interpreted as a classical ensemble of outcomes with certain probabilities. In this case, there are two outcomes (spin up or down), each with probability $1/2$. Recall, the density matrix describing a pure state satisfies ${\rm tr} \, \rho^2 = 1$, which no longer holds here.  

We see that decoherence can, For All Practical Purposes, reproduce the phenomenology of von Neumann projection: there are two possible outcomes, and only one is perceived by a semiclassical observer.

\section{Macroscopic Superposition and Environmental Decoherence}

Decoherence is also responsible for the emergence of a semiclassical reality from the quantum realm, due to interactions with the environment. 

\begin{figure}[htp]
    \centering
    \includegraphics[width=0.6\linewidth]{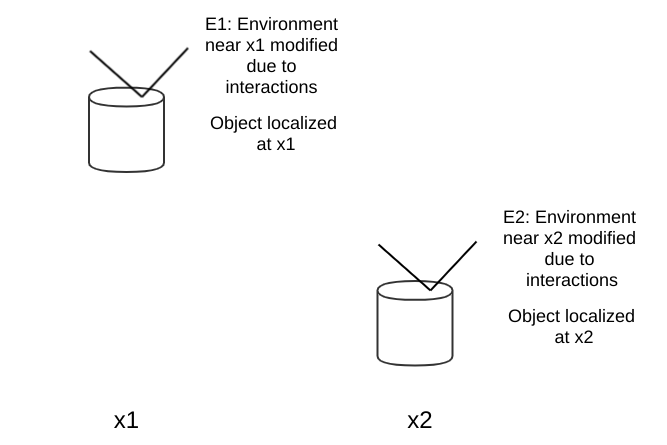}
    \caption{Interactions between the environment (e.g., air molecules) and a macroscopic object localized at $x_1$ lead to a modified environmental state $\ket{E_1}$, with scattered air molecules in the region near $x_1$. If the object is placed at $x_2$ the resulting modified environmental state $\ket{E_2}$ describes scattered air molecules in the region near $x_2$. $\ket{E_1}$ and $\ket{E_2}$ differ macroscopically and have nearly zero overlap.
    }
    \label{fig2}
\end{figure}

Consider a macroscopic object localized at spatial position $x$. Now form a superposition state with support at two different locations $x_1$ and $x_2$ -- i.e., a superposition of the object localized at $x_1$ with another state in which it is localized at $x_2$. What happens when this superposition state interacts with the environment? Assume for example that the environment contains air molecules and photons which scatter from the object. The environment near the object is altered by its presence. Let $E_1$ be the environment state which results from interactions with the object at $x_1$ and $E_2$ be the environment state after interaction with the object at $x_2$. For example, when the object is at $x_1$ the air molecules near $x_1$ scatter from it, but near $x_2$ the air molecules are undisturbed. Placing the object at $x_2$ rather than at $x_1$ reverses this pattern, so clearly $E_1$ and $E_2$, describing the air molecules, are very different environmental states. See figure (\ref{fig2}).

Under (linear) Schr\"{o}dinger evolution, the superposition state evolves as 
\begin{equation}
\frac{1}{\sqrt{2}}
\bigl( \, \vert x_1 \rangle + \vert x_2 \rangle \, \bigr) \otimes \vert E \rangle ~\longrightarrow~
\frac{1}{\sqrt{2}} \bigl( \, \vert x_1 E_1 \rangle + \vert x_2 E_2 \rangle \, \bigr) ~~.
\label{macro}
\end{equation}
We can express the content of (\ref{macro}) as follows. The macroscopic object becomes entangled, through interaction, with the many degrees of freedom in the environment. The two branches of the superposition state 
$~\vert x_1 \rangle + \vert x_2 \rangle~$ decohere from each other due to interactions with the environment $E$, specifically because $E$ evolves into very different states $E_{1,2}$ depending on the location of the object.

The reduced density matrix describing the object position is
\begin{equation}
\rho_x = \frac{1}{2} \Bigl(   \vert x_1 \rangle \langle x_1 \vert  ~+~  \vert x_2 \rangle \langle x_2 \vert
~+~
\langle E_2 \vert E_1 \rangle \,
\vert x_1 \rangle \langle x_2 \vert
~+~
\langle E_1 \vert E_2 \rangle \,
\vert x_2 \rangle \langle x_1 \vert
\Bigr)~.
\end{equation}
Because $E_1$ and $E_2$ have very small overlap, we obtain
\begin{equation}
\rho_x \approx \frac{1}{2} \bigl( \, \vert x_1 \rangle \langle x_1 \vert  ~+~  \vert x_2 \rangle \langle x_2 \vert \, \bigr) ~~.
\end{equation}
This is again a mixed state, not a pure state. At this level of approximation, For All Practical Purposes, an observer either perceives the object at $x_1$ or $x_2$, but not both.

Macroscopic superposition states of this kind can now be created in the lab: for example, drum-like mechanical resonators measuring around 10 microns across have been placed into superpositions of different vibrational modes \cite{Lab1}. The state of the macroscopic drum membrane, at a given instant in time, is a superposition with support in different positions, analogous to the setup analyzed above. To maintain the superposition state (avoid decoherence) the drum system has to be isolated from environmental interactions.

\section{Summary and Further Reading}

The purpose of these notes is to make students aware of the modern perspective on quantum measurement, taking into account decoherence. See \cite{deco1,deco2,deco3} for more comprehensive discussions.

Because of decoherence observers inside the system perceive outcomes consistent with the von Neumann projection postulate even if the system as a whole never deviates from Schr\"{o}dinger time evolution. That is, we can simply assume that the quantum state of the system (observer, measurement device, environment, etc.) evolves according to the Schr\"{o}dinger equation at all times. The projection postulate is not needed for quantum mechanics to describe the experiences of observers inside the system.

What is conventionally referred to as {\it measurement} is actually a continuous process that results from entanglement between the measured object and the many degrees of freedom in the measuring device or local environment. 

Distinct branches of macroscopic superposition states lose contact with each other as they become entangled with environmental degrees of freedom. The branches are said to decohere from each other. 

Under Schr\"{o}dinger evolution, which is unitary, the branches never fully disappear from the theory \cite{deco4}. In fact, isolated quantum systems described by Schr\"{o}dinger evolution can be shown to spend most of their time in macroscopic superposition states \cite{Buniy:2020dux}. Macroscopic superposition states have been realized in the laboratory \cite{Lab1,Lab2} -- there is no experimental evidence against the possibility that you might be in a superposition state as you read this article \cite{deco4}.

The main open problem is to explain the Born rule, which associates probabilities $\vert c_\pm \vert^2$ with outcomes of measurements on superposition states as in (\ref{super}). In traditional formulations of quantum mechanics the Born rule is introduced ad hoc together with the projection postulate. Under pure Schr\"{o}dinger evolution it requires further justification \cite{Hsu:2011cr,Hsu:2015qev}.

\section{Appendix: Density Matrices and Reduced Density Matrices}

A density matrix $\rho$ describes the quantum state of a physical system.
It is a generalization of the state vector $\psi$ which represents a pure state. For a pure state $\rho = \vert \psi \rangle \langle \psi \vert$, and satisfies ${\rm tr} \, \rho^2 = 1$. 

However, density matrices can also represent mixed states. One example of a mixed state arises due to incomplete information: suppose we are told that $\psi$ is randomly selected from an ensemble with equal probability of being in state $\ket{+}$ and $\ket{-}$. Then the density matrix is
\begin{equation}
\rho = \frac{1}{2} \Bigl( \ket{+} \bra{+} ~+~ \ket{-} \bra{-} \Bigr)~~. 
\end{equation}
Note for this mixed state ${\rm tr} \, \rho^2 = 1/2$.

Another situation in which a mixed state arises is in the description of a degree of freedom $A$ which is entangled with other degrees of freedom $B$. While the combined state $\Psi_{AB}$ of $A$ and $B$ may be a pure state, we cannot describe $A$ by itself as such -- it is in a mixed state.

The {\it reduced} density matrix describing $A$ is obtained by tracing over $B$ (i.e., summing over the basis vectors of the $B$ Hilbert space):
\begin{equation}
\rho_A = {\rm tr}_B \, \rho_{AB} = {\rm tr}_B ~ \ket{\Psi_{AB}} \bra{\Psi_{AB}}~~.  
\end{equation}
Expectations of operators $O_A$ acting on the $A$ subspace are given by $\langle O_A \rangle = {\rm tr} \, O_A \, \rho_A$.
Under the assumption of entanglement between $A$ and $B$, $\rho_A$ describes a mixed state. For more details see, e.g., section 2.4.6 of \cite{deco3}.

\section{Acknowledgements}

The author thanks Alexander Adams, Tim Raben, and Lou Lello for useful comments.


\begin{thebibliography}{0}

\bibitem{Bell}
J. S. Bell, Against Measurement, Phys. World 3, 33-40 (1990).
\href{https://docs.google.com/file/d/0ByYDxaP-OyVjVXAtV2VqQlVNMTg}{PDF online.}

\bibitem{deco1}
D. Giulini, C. Kiefer, E. Joos, J. Kupsch, I. O. Stamatescu, H. D. Zeh, Decoherence and the Appearance of a Classical World
in Quantum Theory, Springer (2003).

\bibitem{deco2}
W. H. Zurek, Decoherence, Einselection, and the Quantum Origins of the Classical, Rev. Mod. Phys. 75, 715-775 (2003).

\bibitem{deco3}
M. Schlosshauer, Decoherence and the Quantum to Classical Transition, Springer (2007).

\bibitem{deco4}
One of the earliest treatments of decoherence theory appears in   H. Everett III, The Theory of the Universal Wavefunction (PhD thesis), reprinted in B. S. DeWitt and R. N. Graham
(eds.), The Many-Worlds Interpretation of Quantum Mechanics, Princeton University Press (1973). Everett uses nonstandard terminology such as {\it quantum correlation} in place of {\it entanglement}.

%\cite{Buniy:2020dux}
\bibitem{Buniy:2020dux}
R.~V.~Buniy and S.~D.~H.~Hsu,
Macroscopic Superposition States in Isolated Quantum Systems,
Found. Phys. \textbf{51}, no.4, 85 (2021)
doi:10.1007/s10701-021-00477-2
[arXiv:2011.11661 [quant-ph]].

\bibitem{Lab1}
L. Mercier de L'Epinay et al., Quantum Mechanics-Free Subsystem with Mechanical Oscillators, Science 7 May 2021, Vol 372, Issue 6542, p.625
doi/10.1126/science.abf5389

\bibitem{Lab2}
K. S. Lee et al.,
Entanglement between Superconducting Qubits and a Tardigrade, \href{https://arxiv.org/abs/2112.07978}{https://arxiv.org/abs/2112.07978}.





%Tardigrade superpositions, macroscopic drum heads, etc.


%\cite{Hsu:2011cr}
\bibitem{Hsu:2011cr}
S.~D.~H.~Hsu,
On the Origin of Probability in Quantum Mechanics,
Mod. Phys. Lett. A \textbf{27}, 1230014 (2012)
doi:10.1142/S0217732312300145
[arXiv:1110.0549 [quant-ph]].

%\cite{Hsu:2015qev}
\bibitem{Hsu:2015qev}
S.~D.~H.~Hsu,
The Measure Problem in No-Collapse (Many Worlds) Quantum Mechanics,
Int. J. Mod. Phys. D \textbf{26}, no.03, 1730008 (2016)
doi:10.1142/S0218271817300087
[arXiv:1511.08881 [quant-ph]].





\end{thebibliography}
\end{document}